# Generate Uniform Transverse Distributed Electron Beam along a Beam Line[*]

JIAO Yi (焦毅), CUI Xiaohao(崔小昊)

*Key Laboratory of Particle Acceleration Physics and Technology,*

*Institute of High Energy Physics, Chinese Academy of Sciences, Beijing 100049, P.R. China*

**Abstract:** It has been reported that transverse distribution shaping can help to further enhance the energy extraction efficiency in a terawatt, tapered X-ray free-electron laser. Thus, methods of creating and keeping almost uniform transverse distributed (UTD) beam within undulators are required. This study shows that a UTD electron beam can be generated within evenly distributed drift sections where undulators can be placed, by means of octupoles and particular optics. A concrete design is presented, and numerical simulations are done to verify the proposed method.

Key words: uniform transverse distributed beam, octupoles, particular optics

## 1 Introduction

Modern accelerator physics and applications require various innovative beam manipulation technologies to achieve a satisfactory machine performance, such as reducing bunch length with magnetic compressors [1], modulating beam energy with laser [2], exchanging emittance with deflecting cavity [3], and uniforming transverse distribution with multipoles [4]. Generating uniform transverse distributed (UTD) beam is beneficial in many applications of charge particle irradiation, to release the radiation heating and radiation damage problems, caused by a local bright spots (e.g., in a typical Gaussian beam).

The concept of using third order focusing or even higher order focusing to smooth the nonuniform intensity was first proposed by P.F. Meads [4] and analyzed by others [5-7], and has been verified experimentally [8-9]. The core is to focus the edges of the particle transverse distribution back to the center. To generate a UTD beam, one can either adopt one magnet integrated with multi high-order fields [5-6], or use a pair of octupoles located at particular positions [7-8].

To the best acknowledge of the authors, in previous studies scientists focus on how to generate a UTD beam at a fixed location. Recently it was shown [10] that in a tapered X-ray free electron laser (FEL) with terawatt-level output power [11], if the transverse electron distribution can be flattened from Gaussian to uniform, the extraction efficiency can be largely enhanced, by about 60%. It requires to generate a UTD beam along a beam line, instead of at a fixed point. One may expect that a UTD beam can be produced at the very beginning of an accelerator by shaping the incident laser of a photocathode gun [12]. However, such a distribution may be destroyed by adiabatic damping and various collective effects after acceleration and transportation through a linac accelerator of typically a few hundred meters. Thus, it is best to start the beam uniformization at a position close to where the UTD beam is required. In Ref. [7], a concise way of generating UTD beam at a fixed target by means of a pair of octupoles has been proposed and thoroughly analyzed.

[*] jiaoyi@ihep.ac.cn, supported by National Natural Science Foundation of China (11475202, 11405187) and Youth Innovation Promotion Association of Chinese Academy of Sciences (No. 2015009)

In this paper, the possibility of extending this idea to generate UTD electron beam within evenly distributed drift sections (where undulators can be placed) is discussed in Sec. 2, and a concrete lattice design using typical X-ray FEL beam parameters is presented in Sec. 3. Finally, discussions and conclusive remarks are given in Sec. 4.

**2 Generate UTD beam within evenly distributed drift sections**

It is known that an octupole can be used to apply a transverse kick to particles, with the kick proportional to the cube of the transverse displacement,

$$x = x_0,$$
$$x' = x'_0 - \frac{1}{6}K(x_0^3 - 3x_0 y_0^2),$$
$$y = y_0, \tag{1}$$
$$y' = y'_0 + \frac{1}{6}K(3y_0 x_0^2 - y_0^3),$$

where a thin-lens octupole has been assumed, with $K$ being the octupole integral strength.

Now let us consider a beam line including two octupoles, with the schematic layout shown in Fig. 1. It has been shown [7] that if with a π-section between two octupoles and with a particular ratio between two octupole strengths $K_1/K_2 = -\beta_{x,2}\beta_{y,2}/\beta_{x,1}\beta_{y,1}$ (or more simply, $K_1/K_2 = -1$ with $\beta_{x,1} = \beta_{y,2}$ and $\beta_{x,2} = \beta_{y,1}$, only this case will be considered in below), the transverse coordinates at the observation point after the two octupoles will have only uncoupled, not higher than third-order terms,

$$u_{ob} = a u_0 - b u_0^3 + c u'_0, \tag{2}$$

with

$$a = -\sqrt{\frac{\beta_{u,ob}}{\beta_{u,1}}}(\cos\varphi_u + \alpha_{u,1}\sin\varphi_u),$$
$$b = \frac{K_2}{6}\sqrt{\frac{\beta_{u,ob}}{\beta_{u,1}}}\frac{\beta_{u,2}^2 - \beta_{u,1}^2}{\beta_{u,1}}\sin\varphi_u,$$
$$c = \sqrt{\beta_{u,1}\beta_{u,ob}}\sin\varphi_u,$$

where $u$ stands for $x$ or $y$; the subscripts '0' and 'ob' indicate locations right before the first octupole and of the observation point, respectively; $\beta_{u,i}$ and $\alpha_{u,i}$ ($i = 1, 2, ob$) are the Courant-Snyder (C-S) parameters at the first octupole, the second octupole and the observation point, respectively; and $\varphi_u$ is the phase advance between the second octupole and the observation point.

In Ref. [7], a specific case with up-right phase-space ellipse right before the first octupole is considered. In such a case, $\alpha_{u,1} = 0$, and the ratio between the coefficients '$b$' and '$a$' becomes

$$r = \frac{b}{a} = \frac{K_2(\beta_{u,2}^2 - \beta_{u,1}^2)\varepsilon_u}{6}\tan\varphi_u. \tag{3}$$

In addition, study showed that for an initial Gaussian beam, the final distribution with coordinates in the form of $x = x_0 - kx_0^3$ will be almost uniform when

$$k \equiv r\sigma_{u,0}^2 = \frac{K_2(\beta_{u,2}^2 - \beta_{u,1}^2)\varepsilon_u}{6} \tan\varphi_u \sim 0.1. \tag{4}$$

From this condition, the optimal phase advance $\varphi_{u,opt}$ can be calculated for a fixed $K_2$; and verse visa. It is easy to know the required octupole strength to generate a UTD beam will decrease as $\varphi_u$ changing from $n\pi$ to $n\pi+\pi/2$ ($n$ is integer). However, the contribution from the distribution in $x'_0$ (typically in Gaussian) will increase at the meantime [cf. Eq. (2)], causing less uniform profile. Thus, it was recommended to choose $\varphi_u$ close to $n\pi$, with certainly a price of relatively high octupole strength.

We note that the condition (4) also applies to the case with a tilted phase-space ellipse right before the first octupole. To increase the focusing efficiency, it is generally to increase the beam size at the octupole, resulting in an elongated ellipse in phase space. The relation, $u'_0/u_0 = -\alpha_{u,1}/\beta_{u,1}$, holds approximately, with the ratio $-\alpha_{u,1}/\beta_{u,1}$ representing the tilt of the phase space ellipse. With this relation, Eq. (2) can be simplified to

$$u_{ob} \approx \tilde{a}(u_0 - ru_0^3), \tag{5}$$

with $r$ in the same expression as in Eq. (3). With a similar derivation as in Ref. [7], one can obtain the same uniformization condition as in Eq. (4).

Of particular interest, the beam density uniformization at an unfixed observation point along a beam line will be discussed in the following.

First, since $\tan\varphi_u$ is periodic in $\pi$, uniform distribution will appear after a $\pi$-section [cf. Eq. (4)]. And the second, it is empirically found that when $k$ is slightly varied around 0.1 (e.g., $\Delta k = +/-0.03$), the uniform transverse profile can be almost kept, indicating that it is possible to kept a UTD beam along a drift section (where an undulator can be placed) with $\varphi_u$ at the center of the section being $\varphi_{u,opt} + n\pi$. The dependency of $k$ on $\varphi_u$ can be derived from condition (4),

$$dk \approx \frac{0.2}{\sin(2\varphi_{u,opt})} d\varphi_u. \tag{6}$$

Thus, a reasonable estimation of the available variation range of $\varphi_u$ is $\Delta\varphi_{u,avaiv} \approx \pm 0.15\sin(2\varphi_{u,opt})$. The largest $\Delta\varphi_{u,avaiv} \approx \pm 0.15$ occurs at $\varphi_{u,opt} = \pi/4+n\pi/2$. Considering a drift section where the beta function $\beta_{drift}$ remains approximately constant (an assumption usually adopted in FEL theory), the available maximum length of the drift section where a UTD beam is kept can be estimated by

$$L_{drift,avaiv} \approx 0.3\beta_{drift} \sin(2\varphi_{u,opt}), \tag{7}$$

where the relation between phase advance and beta function, $\Delta\varphi = \int ds/\beta$, has been used.

All the above considered, one can generate a UTD beam within evenly distributed drift sections along a beam line, with a $\pi$-section between adjacent drift sections; moreover, to preserve the uniformity as long as possible, it is best to choose $\varphi_u$ at the center of the drift section to be $n\pi-\pi/4 < \varphi_u < n\pi$, or $n\pi < \varphi_u < n\pi+\pi/4$.

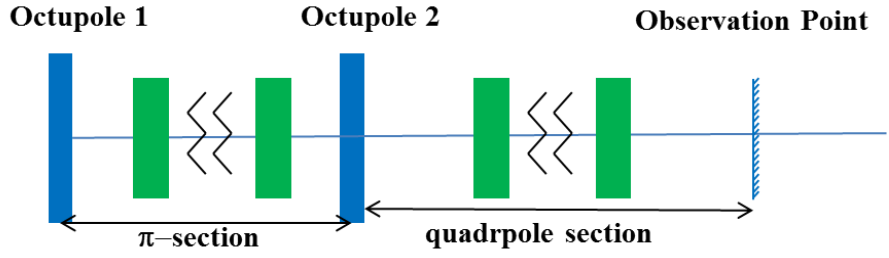

Fig. 1. Schematic layout of the transport system, where two octupoles with a π-section in between are used to generate a UTD beam at the observation point.

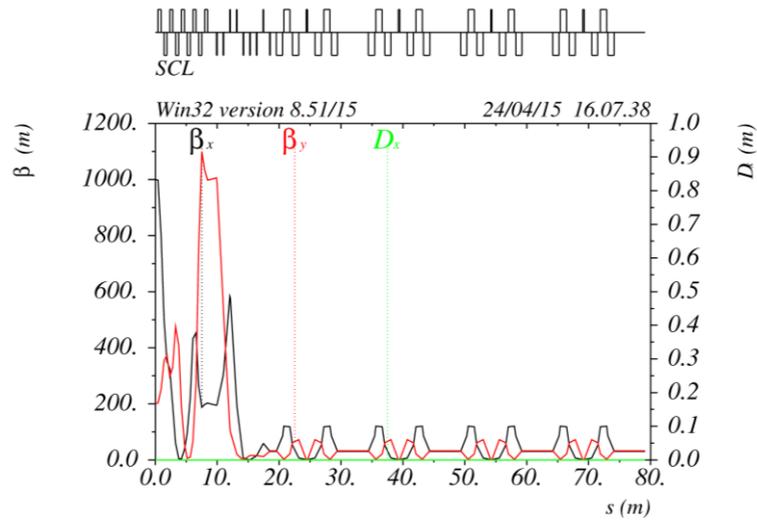

Fig. 2. Optical functions along the beam line with octupoles (located at $s = 0$ and 8.9 $m$) and with evenly distributed 5-m drift sections.

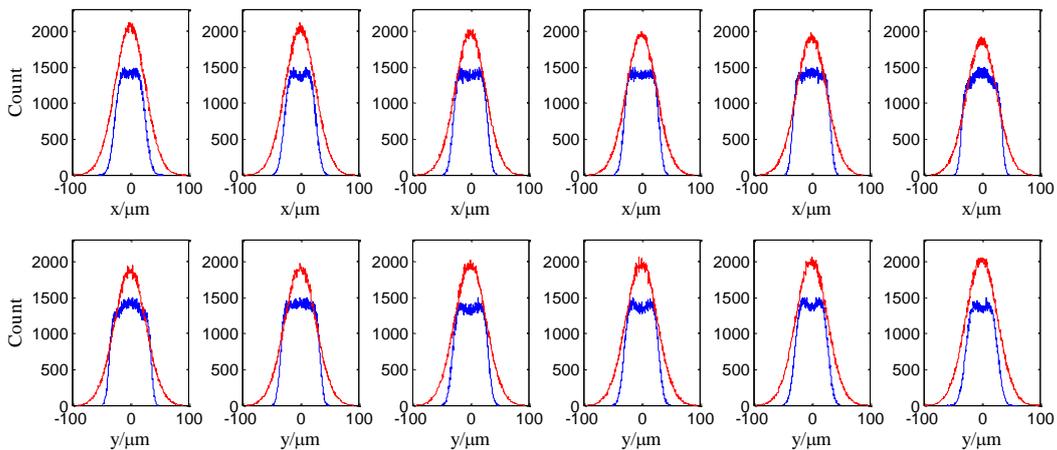

Fig. 3. Beam distribution profiles in $x$ (upper plots) and $y$ (lower plots) planes along the 5-m drift section with octupoles turned on (blue curves) and off (red curves), obtained with ELEGANT simulation. In each row, the six plots from left to right stand for different positions of the drift section with distance of 1 $m$.

## 3 Concrete design of a beam line generating a UTD beam

Based on the above analysis, a concrete design of a beam line generating UTD beam within evenly distributed drift sections has been made and will be presented below.

As we know, in X-ray FELs electron beam has intrinsically ultralow emittance. Here a particular case with LCLS-like beam parameters [11] with normalized emittance of 0.3 mm.mrad and beam energy of 13.64 GeV is considered. Very strong octupole focusing is required for beam intensity uniformization. Enlarging the beta function at octupole is likely to have only limited effect on reducing the octupole strength (the beam size that proportional to $\beta^{1/2}$ should be much smaller than the vacuum chamber). For instance, with $\beta_{x1}$ = 1000 m and $\beta_{x2}$ = 200 m, $\varphi_x$ = $\pi/10+n\pi$ at the center of the drift section, from Eq. (4), the required $K_2$ is about $1.7 \times 10^5$ m$^{-3}$. If with normal magnetic design technology, impractically thick octupoles are required. Fortunately, recent progress in vacuum and magnet technology allows one to achieve high gradient by adopting small-aperture magnets. To illustrate, if with the bore radius of 10 mm and with pole face filed of 0.8 T, the required focusing can be realized with an octupole of 1.6 m (or shorter if using a superconducting octupole). The magnet aperture size is very small, but is still larger than 90 folds of the beam size (~ 106 μm).

Experience [10, 11] showed that for a terawatt X-ray FEL with LCLS-like parameters, the optimal beam size within undulators is about 18 μm, corresponding to an average beta function of 29 m. From Eq. (7), $L_{drift,avaiv} \approx$ 5.1 m for $\varphi_{x,\ opt}$ = $\pi/10+n\pi$. On the other hand, owing to the limitation of technological level at present, one undulator magnet is typically not longer than 5 m. Thus, it appears feasible to keep a UTD beam within the undulators.

Another important concern is to minimize the length of the π-section between undulators. To this end, a section consisting of quadrupoles and drifts is considered. All the quadrupole strengths and the drift lengths are treated as variables and optimization is performed in multi-variable space. The result shows that to realize phase advance of π in both $x$ and $y$ planes, at least seven quadrupoles and enough drift space between quadrupoles (not smaller than 5 m) are required.

Combining the above considerations, a concrete design has been made. As shown in Fig. 2, after two octupoles, drift sections of 5 m are evenly distributed along the beam line. The beam line is then modeled with the ELEGANT code [13] and particle tracking is performed for the cases with octupoles turned on and off, respectively. The results are presented in Fig. 3. One sees that uniform transverse distribution remains within the 5-m drift section when turning on the octupoles, which agrees very well with the analytical prediction.

## 4 Discussions

As inspired by the advance in terawatt X-ray FEL research, in this paper we explore the feasibility of the density uniformization of electron beam in transverse planes, instead of a charged hadron beam as in previous studies. It is illustrated that a UTD electron beam can be generated and kept within undulators in a terawatt X-ray FEL by using a pair of octupoles and particular optics design. A concrete design for the particular case with LCLS-like parameters is presented and the method is verified with numerical simulations.

Since electron beam usually has ultralow emittance in FELs, which causes a few difficulties in an actual implementation of the intensity uniformization using octupoles. First, it requires large beta function to create a moderate beam size at octupole, and meanwhile, needs very strong octupole focusing. In the presented design for the particular case with LCLS-like parameters, an octupole with bore radius as small as 10 mm is required. It is feasible, but should be a great challenge, to fabricate such a small-aperture octupole. Secondly, to achieve uniform distribution in undulators, it calls for a π-section between adjacent undulators; and for a promising FEL performance, the π-section should be as short as possible. However, to realize a phase advance of π in both $x$ and $y$ planes, it needs strong quadrupole focusing and enough drift space between quadrupoles, leading to a longer distance between undulators than usual (e.g., the π-section is of about 10 m in the presented design). This will inevitably increase the total length of the undulator section, and cause additional radiation slippage effect (e.g., for X-ray wavelength of $\lambda_r = 1.5$ Å, the slippage over a 10-m section is about $47\lambda_r$). As a result, an electron beam longer than usual will probably be required to ensure the electron-radiation interaction over the whole undulator section.

We note that these difficulties can be alleviated, in some degree, if adopting electron beam with slightly larger normalized emittance (e.g., 1 mm.mrad) and lower energy (e.g., 8 GeV). Keeping other parameters unchanged, the required octupole length will decrease by 80%, and the π-section between undulators will be shorter (~ 8 m). Besides, the slippage effect will be weaker for a longer FEL wavelength (e.g., soft X-ray range). Regardless, we admit further in-depth studies are necessary to make the design more sophisticated and more practical. Evaluation of the FEL performance with such a design is also under way and will be presented in a forthcoming paper.

Finally, we would like to point out that such a design philosophy, i.e., generating UTD beam within evenly distributed drifts, may find its possible applications in hadron machines, where the octupole strength and the length of the π-section may not become a great concern.